# Drishtikon: An advanced navigational aid system for visually impaired people

Shashank Kotyan, Pankaj Kumar Sahu, Nishant Kumar, Venkanna U.

*Abstract*—**Today, many of the aid systems deployed for visually impaired people are mostly made for a single purpose. Be it navigation, object detection, or distance perceiving. Also, most of the deployed aid systems use indoor navigation which requires a pre-knowledge of the environment. These aid systems often fail to help visually impaired people in the unfamiliar scenario. In this paper, we propose an aid system developed using object detection and depth perceivement to navigate a person without dashing into an object. The prototype developed detects 90 different types of objects and compute their distances from the user. We also, implemented a navigation feature to get input from the user about the target destination and hence, navigate the impaired person to his/her destination using Google Directions API. With this system, we built a multi-feature, high accuracy navigational aid system which can be deployed in the wild and help the visually impaired people in their daily life by navigating them effortlessly to their desired destination.**

## I. INTRODUCTION

World Health Organisation (WHO) estimates that 253 million people worldwide are severely or moderately visually impaired [1]. Often visually impaired people are aided by the traditional way of using a cane to navigate and find their way. This method fails to aid visually impaired people when they fail to find the appropriate markers or they realise too late. Also, this method of using canes fail during the anxiety attack or when navigating in an unexplored environment. Some visually impaired people use dogs to guide and navigate their way, but this requires a lot of time as the dog requires training. As one may notice, that these methods do not benefit from the boon of technology. Today, as the science has improved and achieved great heights there still lies a group of people who are yet to be benefitted by the technological achievements. There still lies a great potential in the field of healthcare and aiding systems which are yet to be explored and benefitted by the technology.

Computer vision (CV) is one of the many branches of computer science which is thriving to resemble human vision. It is not confined to capture an image from the camera as it also encapsulates an intelligent software module which is used for analysing the image based on various algorithms for specific applications. These algorithms incorporate processing of image/s to determine the features and attributes which can describe the image at an abstract level. These features and attributes of an image provide meaningful inferences with respect to specific applications. This involvement of intelligent software module classifies CV as a category of science and it is growing to become a perfect artificial vision [2]. However, computer vision is also trying to teach a system to pick what should be a zone of interest, how to understand the objects and how the machine should be thinking. The applications of CV has grown to almost all the domains of computer science which involves analysis and interpretation of data and image, since the past 30 years. Along with applications in computer science domain, CV has found its application in various interdisciplinary fields [3]. Some of the applications of computer vision are monitoring the industrial equipment, monitoring biological cells and some real-life applications like parking systems. The technological implementation of CV system is better than biological in many ways like a camera can detect movements faster than the human eye. Due to this reason, CV has also found its deployment in the area of healthcare for visually impaired people.

The average lifespan of the humans has increased so as the number of elderly people which makes healthcare as one of the most important social issues in a human life. This demographic shift has increased the burden on healthcare services. With this ever-increasing problem, there has been a requirement of various aid systems for the people to decrease the burden on healthcare services. Some of the notable examples of aid systems are auditory aid systems and fall detection systems which aid the elderly people in their daily routine. There are also specific aid systems like sleep monitoring and action monitoring systems which help the medical professionals study and monitor their subjects without any hindrance. These technology-focused aid systems have been in the research for long but very few deployable and usable aid systems exist in society for helping those who are in need. While doing the social services in the university, we encountered a group of visually impaired people who were dependent on their guides for navigating them in an unfamiliar environment, while other physically disabled people rarely required any assistance for communicating or navigating. The visually impaired people faced a lot of trouble doing the normal activities.

Navigation systems are one type of aid systems which help a person navigate in an unfamiliar environment smoothly without being lost or getting hurt. In an indoor navigation system, the system helps to navigate a person in an indoor controlled environment where the environment is usually fully observable. The main problem which an indoor navigation system faces is of the lack of precision which is due to the localization method used. There are various methods which are popularly used are pre-installed indoor communication infrastructures, laser, radar, sonar, camera, motion sensors, etc. In an outdoor navigation system, the system helps to navigate a person in the outdoor uncontrolled environment. The various methods popularly used are GNSS, field strength measurements using either WLAN, GSM or Bluetooth. Also, some of the global and local positioning systems are used such as Loran and VOR-DME. As we mentioned before, that computer vision aims to resemble human vision, therefore, it can be used as a substitute for the people who are devoid of such sense.

## II. RELATED WORKS

Present aid systems deployed for visually impaired people involved a mix of computer vision and other technologies with various limitations.

Mekhalfi et.al. proposed a sensing approach to describe indoor scenes with the help of a portable camera. The major limitation of the prototype was scalability to adapt for more objects and high processing power which makes the system bulky and high power consuming [4]. Hub et.al. proposed an indoor navigation system with the object identification feature using the colour of the objects. It makes use of enhanced processing power of mobile devices to allow real-time processing of 3D models and local sensors of orientation and object identification. The major flaw of the prototype was inaccurate depth measurement which can cause fatal results if deployed in the real world [5]. Kumar et.al. proposed RFID and GPS integrated navigation system for visually impaired which required a pre-developed infrastructure of RFID tags. It also used sensors such as ultrasonic and infrared to identify the objects works in the very close vicinity and will require the user to be physically near the object in order to detect it which makes it infeasible in the real-life navigation systems as the visually impaired person can dash into the object before detecting it [6]. Kaiser et.al. developed a wearable navigation system for the visually impaired which uses the SLAM approach for navigation in the indoor environment and requires infrastructure to work in an open indoor environment. The purpose of the system was to navigate a person by localizing its location every time the user makes a step. Another notable limitation of the indoor navigation system is, they require either developed infrastructure or sensor fusion which drastically increases the cost of the navigation system. Also, these systems are specialized for the specific indoor environment and therefore, cannot be readily used in an unfamiliar environment [7]. The outdoor navigation systems, methods help to navigate a person in long distances but fail to accurately help the person navigate in close vicinity.

Neto et.al. proposed a Kinect-based wearable real-time face recognition system with audio feedback. They developed a high-accuracy face recognition system by taking advantage of depth information and temporal coherence which is the simple and computationally efficient approach. This method has the limitation of limited scope and usable only in close vicinity which makes them efficient in an indoor environment but inefficient in the outdoor environment [8]. Dakopoulos et.al. reviewed a set of wearable travel aid systems for avoiding an obstacle in which most of the object detection and distance calculation were done by the outputs of sensors such as SONAR, Infrared and Ultrasonic sensors. Few of the systems also employed a tactile support instead of the auditory response which makes the usability of the system a bit tough due to the training required for learning the tactile response of the systems [9]. Takizawa et.al. proposed a cane-based method for assisting blind people which aids the user similar to a traditional cane but it requires excessive human interaction to work and has a huge size which makes it an impractical solution for aiding visually impaired people [10]. Another limitation of the use of cane-based systems is that they can be only used for detecting and identifying the objects in the close vicinity. These systems only detect the objects when the cane is touched upon the object and works like the traditional cane with the benefit of telling the object to the user. Sometimes, these learning were frustrating to the user and therefore, are an inefficient means to communicate with the user.

The main limitation of the current aid systems is the bulky nature of the system which is not feasible for the elderly which constitute a large portion of the visually impaired society. The system must be as light as possible and easy to use without much technology related knowledge as pre-requisite. The main advantage of computer vision systems over other techniques is the absence of a requirement of an infrastructure as computer vision systems can be mounted over existing visual systems which make them much scalable than the sensor networks. Also, the sensors have a relatively less range for object detection than the computer vision systems. As the systems employing computer vision uses a processing module and a camera, it is usually less bulky then the counterparts developed using sensor networks. Another major advantage of using machine vision systems over other techniques is the replacement which vision systems have did to sensor networks in few domains such as parking system due to their high scalability. Some of the common limitations which are addressed in our solution are a smart navigation system for not only navigating a person from one location to another but at the same time preventing the person from dashing into objects in close vicinity which can hurt the person.

## III. DRISHTIKON

Our proposed solution can be classified into three independent working modules as shown in Figure 1 which are explained further in the section. Table 1 gives the hardware and power supply requirements. Table 2 gives the software, libraries and algorithm requirements of the proposed prototype.

### A. Input Module

The input module consists of a pair of Quantum USB Web Camera and a smartphone. Quantum USB Web Camera is used for collecting real-time frames from the environment. The smartphone is used for interacting with the user. The smartphone's microphone is used to listen to the destination address, the user wants to go.

### B. Processing Module:

The processing module consists of a Raspberry Pi (RPi) and a smartphone. RPi is used for receiving input from Web Camera. This input is then used for detecting objects and distance of the object from the user. The result is then sent to the smartphone for communicating the information to the user. After listening to the user for the destination, the smartphone finds the shortest route to the destination from the current

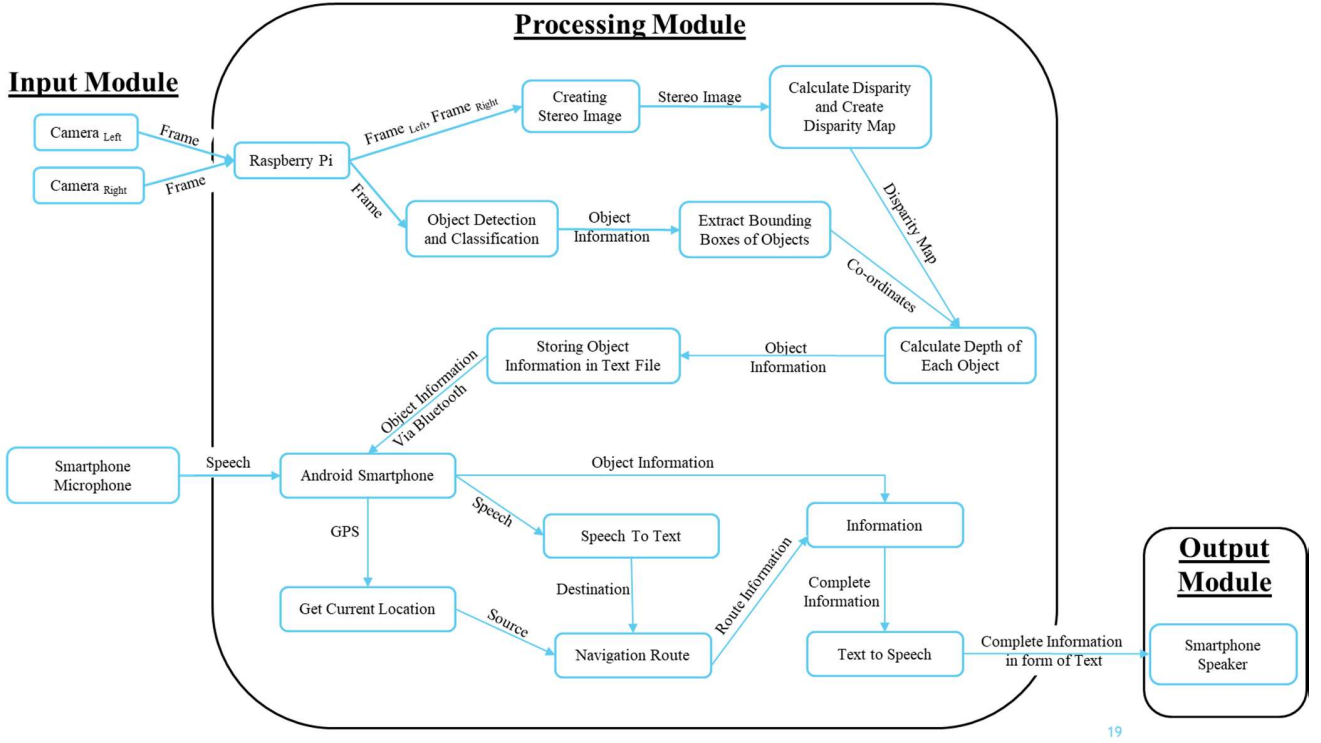

Fig. 1. System Design

TABLE I
HARDWARE REQUIREMENTS

| Task | Component |
|------|-----------|
| *Processing* | Raspberry Pi 2 or higher versions (Recommended: Raspberry Pi 3 Model B) |
| *Image Capturing* | Raspberry Pi Compatible Camera Example, Raspberry Pi Camera, USB Web Camera |
| *Power Supply* | 5V/ 2A through Micro-USB port |
| *Smartphone* | Smartphone with microphone, speaker, internet connection, Bluetooth |

TABLE II
SOFTWARE REQUIREMENTS

| Task | Component |
|------|-----------|
| *Stereo Calibration and Disparity Map* | Open CV Stereo Camera Calibration, Camera Calibration Algorithm. |
| *Navigation* | Google Directions API |
| *Object Detection and Classification* | Tensorflow |
| *Speech to Text and Text to Speech* | Google Voice API |

location.

*a)    Algorithms used in the prototype*

*Object detection and classification:* We used the tensorflow's standard object detection and classification algorithm to detect and localize the object in the frame.

*Stereo camera calibration:* We have used the standard Open CV's algorithm for calibrating cameras for stereo imagery. We have calibrated the cameras using checkerboard which is further described in detail in section 4.

*Creation of disparity map:* Using the images captured by the pair of cameras, we computed the disparity map of the given frames using standard Open CV's algorithm for calculating disparity map.

*Calculating depth from disparity:* We calculated the depth of an object using the standard formula of the depth of a pixel,

$$Depth\ (D) = \frac{Base\ offset\ (b) * Focal\ length\ of\ the\ camera\ (f)}{Disparity\ value\ (d) * Pixel\ size\ (px)} \quad (1)$$

since the formula (1) calculates the depth of a single pixel from the camera, we have used mean to calculate the overall distance of the object,

$$Distance\ of\ Object\ (dist) = \frac{1}{N} \sum_{X=x}^{x+w} \sum_{Y=y}^{y+h} D_{X,Y} \quad (2)$$

Where, N is the total number of pixels enclosed by the object, for simplicity we assume the object will have a rectangular boundary due to which N can be replaced by N=w*h where w and h are width and height of the rectangular boundary enclosing the object. (X, Y) are the order of pixels in 2D space. (x,y) are the order of pixels of the top-left pixel of the object. (x+w, y+h) are the order of pixels of the bottom-right pixel of the object. $D_{X,Y}$ is the depth of the pixel having an order (X, Y)

*Sending a message over Bluetooth:* We have used the standard socket level protocol to communicate between the two Bluetooth devices. We configured RPi to act as a server capable of only sending data to the required client, which in our case is the paired Android device.

*Receiving input from the user in form of speech:* We used to smartphone's microphone to record the audio of the user speaking the desired target destination address.

*Converting speech to text using Google Voice API:* We have used the Google Voice API to convert the recorded speech of target destination into text.

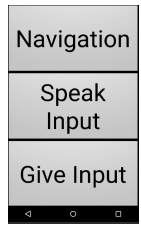

Fig. 2. GUI of Mobile Application

*Finding route between two locations using Google Directions API:* We have used the online Google Directions API to find the best possible route between the current location and target destination.

*Receiving a message over Bluetooth:* We have used the standard socket level protocol to communicate between the two Bluetooth devices. We configured Smartphone to act as a client capable of only receiving data from the server, which in our case is the paired raspberry pi device.

*Converting Text to Speech using Google Voice API:* We have used the Google Voice API to convert the text which is our result into speech. Google Voice API was used because it comes pre-compiled for text to speech for Android versions 4.2 (Jelly Bean v2) and above.

### b) Smartphone Application

The smartphone application was developed keeping in mind the target audience which is visually impaired people. For the simplicity and easy to use, it's UI consists of only three buttons as shown in Figure 2. The bottom or Give Input button listen to the user for a destination. The top or Navigation button retrieves the information from the Google Directions API if the destination given by the user is correct or else sends a beep message to warn the user that the input given was incorrect. The middle or Speak Input button retrieves the information from the Raspberry Pi and then the information is sent to the user via a microphone in a pre-formatted output. A sample of the output is mentioned below,

*Head <Navigation Instruction> but beware there is < <Object Label> is at <Depth of the Object> feet (Repeats till all object's information is passed to the user>.*

### c) Methodology

Our solution's processing is divided into two main modules of RPi and smartphone. RPi detects and identifies an object along with its distance. The result is then sent to a smartphone which interacts with the user and also finds the optimal route between two locations.

The two cameras first capture the environment in the frame and send it to RPi for processing. The RPi uses the frame to compute the stereo image which is used for creating a disparity map of the stereo image. At the same time, it also predicts the visible objects in the frame and captures its location in the frame. The location of the object is then passed to the disparity map to compute the mean distance of the object from the user. This information about the type of object and distance of the object is then passed onto the smartphone for communicating

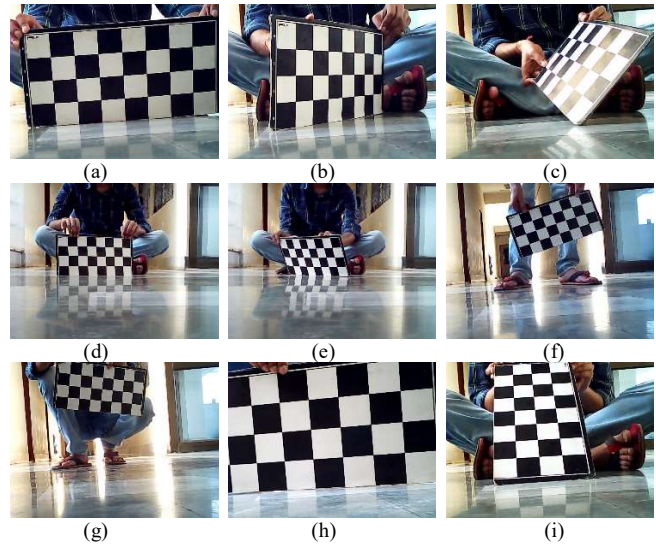

Fig. 3. Strero Calibration Checkboard Images

to the user.

The smartphone receives a speech input from the user which is the target destination of the user for navigation. After receiving the input, the smartphone converts the speech to text using the online Google Voice API. At the same time, smartphone computes the current location of the smartphone for getting the source location. Now, this source and destination information are passed onto Google Directions API for computing the optimal path between the locations. The information, containing objects in nearby vicinity and navigation is passed onto the Text to Speech engine to convert the text into an audio signal. This audio signal is then played from the smartphone's speaker which communicates the required information to the user.

### C. Output Module:

The output module consists of the smartphone speaker which is used to convey the information processed to the user.

## IV. Experiments Conducted

### A. Experiment for stereo camera calibration

We calibrated the Quantum Web Cameras to act as a stereo camera with the help of checkboard of size 8x4. We used the checkboard to detect the sub-checkboard of size 5x4. Some of the images we used in calibrating the stereo camera are shown in Figure 3.

In order to get a good stereo calibration the checkerboard was moved around the camera frame such that the following calibrations were performed:

X calibration: The checkerboard was detected at the left and right edges of the field of view.

Y calibration: The checkerboard was detected at the top and bottom edges of the field of view.

Skew calibration: The checkerboard was detected at various angles to the camera.

Size Calibration: The checkerboard fills the entire field of view and the checkboard is at distance with the camera.

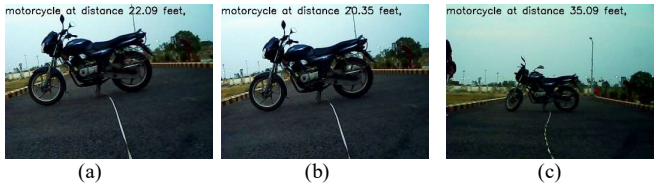

Fig. 4. Detection and depth prediction of Motorcycle

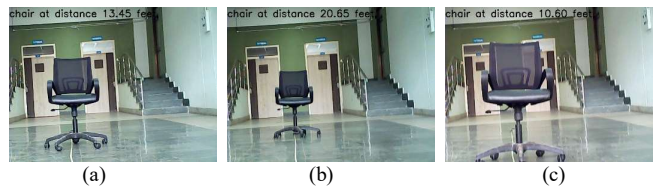

Fig. 6. Detection and depth prediction of Chair

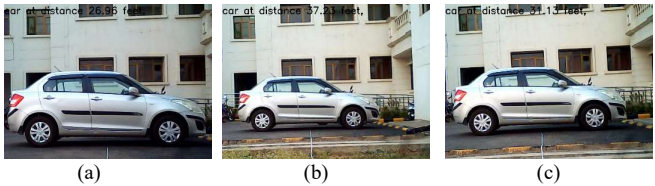

Fig. 5. Detection and depth prediction of Car

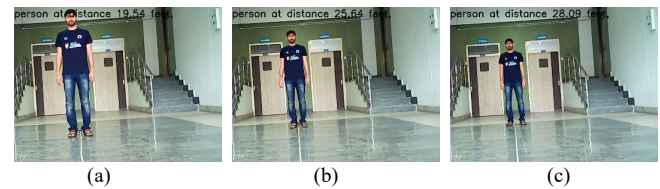

Fig. 7. Detection and depth prediction of Chair

Overall calibration: Checkerboard tilted to the left, right, top and bottom.

### B. Experiment for navigating a visually impaired person in the wild

We conducted a controlled experiment of navigating two volunteers without using their visual senses and solely relying on our solution to navigate to their destination without dashing into an object. The experiment consisted of three phases,

#### 1) Retrieving destination location from the user

In this phase, the user gives the input to the smartphone about the desired destination in the form of voice command.

#### 2) Retrieving information to be given to the user

In this phase, the smartphone gathers the information from the RPi about the objects in close vicinity and the navigation command from the Google Directions API. If either of the information cannot be gathered due to connection error with RPi or wrong destination given from the user, a warning beep is sounded notifying the error to the user.

#### 3) Speaking the information to the user

In this phase, the gathered information is said out loud to the user using the speaker of the smartphone. The speed of the command is adjusted to a slower than the normal human speaking rate to ensure that the user listens to the command efficiently.

As the experiment sent an audio signal to the volunteers which cannot be reproduced in print. Therefore, we present some of the captured frames and the result processed about the objects in context written in the frame as shown in Figure 4-7.

### V. RESULT ANALYSIS

Our proposed solution can successfully and accurately detect and classify objects based on objects which can be dashed into by visually impaired people. With a careful observation, we can see the detailed execution time as shown in Table 3 which lies in the acceptable range.

The proposed solution can accurately classify and detect objects. The classification can be done for 90 different objects which can be perceived in an outdoor environment. Further after modifying the formulae to calculate the focal length of the camera, the disparity map calculated the focal length of 1.9333 while in reality, the focal length of the camera as specified by the manufacturer was 2.0 which rounds of the error of focal length to be -0.0667. This error is because of rms stereo calibration error of 10.257778. The proposed solution has, therefore, an error of calculating distance up to ± 2ft of the objects in the close vicinity.

TABLE III
EXECUTION TIME FOR VARIOUS TASKS

| Task | (Average) Execution Time (in seconds) |
|------|--------------------------------------|
| *Setup the modified HAUAR and other preliminary tasks* | 12 |
| *Setup Disparity Map and Calculating Depth* | 0.6 |
| *Capturing of Images* | 1 |
| *Classifying and detecting object* | 10 |
| *Cumulative time to run one instance of solution* | 12 |

### VI. CONCLUSION

With the result analysis and the comparative analysis of the proposed solution with other implemented solutions, we can safely conclude that our solution primarily focuses on preventing the user to dash into the objects in close vicinity and at the same time provide the optimal navigation to the target destination. We have successfully deployed a non-human intervened system which is capable of taking detecting and classifying objects in close vicinity and find the closest path to the target destination. Our prototype is capable of detecting and classifying 90 objects. This project has been successfully tested in real time with multiple objects in close vicinity and target destination in the walkable distance. The current model is apt for only outdoor navigation as it depends on Google Navigation API to navigate. A possible extension can be to work on indoor navigation which is a standard problem on its own. Also, the current model is composed of two hardware units of an RPi and a smartphone. Another possible extension can be to merge the processing in two hardware units into a single hardware unit for better feasibility and usability.

### REFERENCES

[1] World health organization, media centre, visual impairment and blindness, fact sheet no 282 http://www.who.int/en/news-room/fact-sheets/detail/blindness-and-visual-impairment

[2] Zubaľ, M., Lojka, T., & Zolotová, I. (2016, January). IoT gateway and industrial safety with computer vision. In Applied Machine Intelligence and Informatics (SAMI), 2016 IEEE 14th International Symposium on (pp. 183-186). IEEE.

[3] Walsh, R., & Hornsby, A. (2011, January). Towards off-the-shelf computer vision for user interaction in consumer homes. In Consumer Electronics (ICCE), 2011 IEEE International Conference on (pp. 121-122). IEEE.

[4] Mekhalfi, M. L., Melgani, F., Bazi, Y., & Alajlan, N. (2015). A compressive sensing approach to describe indoor scenes for blind people. IEEE Transactions on Circuits and Systems for Video Technology, 25(7), 1246-1257.

[5] Hub, A., Diepstraten, J., & Ertl, T. (2004, October). Design and development of an indoor navigation and object identification system for the blind. In ACM Sigaccess Accessibility and Computing (No. 77-78, pp. 147-152). ACM.

[6] Yelamarthi, K., Haas, D., Nielsen, D., & Mothersell, S. (2010, August). RFID and GPS integrated navigation system for the visually impaired. In Circuits and Systems (MWSCAS), 2010 53rd IEEE International Midwest Symposium on (pp. 1149-1152). IEEE.

[7] Kaiser, E. B., & Lawo, M. (2012, May). Wearable navigation system for the visually impaired and blind people. In Computer and Information Science (ICIS), 2012 IEEE/ACIS 11th International Conference on (pp. 230-233). IEEE.

[8] Neto, L. B., Grijalva, F., Maike, V. R. M. L., Martini, L. C., Florencio, D., Baranauskas, M. C. C., ... & Goldenstein, S. (2017). A Kinect-based wearable face recognition system to aid visually impaired users. IEEE Transactions on Human-Machine Systems, 47(1), 52-64.

[9] Dakopoulos, D., & Bourbakis, N. G. (2010). Wearable obstacle avoidance electronic travel aids for blind: a survey. IEEE Transactions on Systems, Man, and Cybernetics, Part C (Applications and Reviews), 40(1), 25-35.

[10] Takizawa, H., Yamaguchi, S., Aoyagi, M., Ezaki, N., & Mizuno, S. (2015). Kinect cane: An assistive system for the visually impaired based on the concept of object recognition aid. Personal and Ubiquitous Computing, 19(5-6), 955-965.